# Electronic orders in the Verwey structure of magnetite

Mark S. Senn,[1] Ingo Loa,[2] Jon P. Wright,[3] and J. Paul Attfield[1*]

[1] *Centre for Science at Extreme Conditions and School of Chemistry, University of Edinburgh, West Mains Road, Edinburgh, EH9 3JZ, United Kingdom;*

[2] *SUPA, Centre for Science at Extreme Conditions and School of Physics and Astronomy, University of Edinburgh, Mayfield Road, Edinburgh, EH9 3JZ, United Kingdom;*

[3] *European Synchrotron Radiation Facility, 6 rue Jules Horowitz, Grenoble Cedex 9, 38000 France*

Electronic structure calculations of the Verwey ground state of magnetite, $Fe_3O_4$, using density functional theory with treatment of on-site Coulomb interactions (DFT+U scheme) are reported. These calculations use the recently-published experimental crystal structure coordinates for magnetite in the monoclinic space group *Cc*. The computed density distribution for minority spin electron states close to the Fermi level demonstrates that charge order and $Fe^{2+}$-orbital order are present at the *B*-type lattice sites to a first-approximation. However, $Fe^{2+}/Fe^{3+}$ charge differences are diminished through weak bonding interactions of the $Fe^{2+}$-states to specific pairs of neighboring iron sites that create linear, three-*B*-atom trimeron units that may be regarded as 'orbital molecules'. Trimerons are ordered evenly along most Fe atom chains in the Verwey structure, but more complex interactions are observed within one chain.

## Introduction

Magnetite ($Fe_3O_4$) is the eponymous magnetic substance and occurs widely on earth as a mineral and biomineral. In this $AB_2O_4$ spinel-type material, there are twice as many *B*-site $Fe^{3+}$ $3d^5$ $S = 5/2$ up-spins as there are down-spins at the *A* sites, resulting in a net magnetization. Rapid hopping of an 'extra' down-spin electron between *B* sites, as represented in the formal charge distribution $Fe^{3+}[Fe^{2.5+}]_2O_4$, results in minority-spin-polarised electronic conductivity. On cooling, a sharp first-order transition is observed in measurements of heat capacity, conductivity, magnetisation and many other properties at around 125 K [1,2]. This is accompanied by a complex lattice distortion to a monoclinic √2 x √2 x 2 superstructure of the cubic room temperature lattice [3,4]. The supercell has *Cc* space group symmetry and contains 224 atoms. Verwey proposed that the transition is driven by a regular condensation of $Fe^{2+}$ and $Fe^{3+}$ ions equivalent to localisation of the minority spin 'extra' electrons [1], a phenomenon now known as 'charge ordering' that has been verified in many other oxides [5]. However, the ground state of magnetite has remained a contentious issue for over 70 years, as microtwinning of *Cc* domains below the Verwey transition hampers diffraction studies of the low temperature structure. Partial structure refinements from powder diffraction data [6,7,8] and resonant X-ray studies [9,10,11,12,13] have led to a variety of proposed charge-ordered and bond-dimerised ground state models in recent years [14,15,16,17,18,19,20].

An X-ray refinement of the full low temperature *Cc* superstructure of magnetite was recently reported [21]. A large number of frozen phonon modes were found to contribute to the overall structural distortion. $Fe^{2+}/Fe^{3+}$ charge ordering and $Fe^{2+}$ orbital ordering was proposed from analysis of the observed interatomic distances, showing that Verwey's hypothesis[1] is correct to a useful first approximation. The observed charge ordering pattern is consistent with predictions of two recent electronic band structure calculations based on density functional theory (DFT) with treatment of on-site Coulomb interactions (DFT+U scheme) [16,18], and the observed orbital order was also predicted in ref. 18. However, additional structural distortions in which *B* site Fe-Fe distances within linear Fe-Fe-Fe units are anomalously shortened suggested that the 'extra' down-spin electrons are not fully localised as $Fe^{2+}$ states, but are instead spread over three sites resulting in highly structured three-site polarons termed 'trimerons'. Some $Fe^{3+}$ displacements consistent with trimeron formation were noted in ref. 18, but the apparent dominance of trimeron order within the Verwey structure of magnetite was not proposed in previous theoretical or experimental studies. Here we report electronic structure calculations for the experimental *Cc* structural model, which provide further insights into the charge, orbital and trimeron orders in magnetite.

**Electronic Structure Calculations**

Electronic structure calculations were performed in the framework of density functional theory (DFT) with the full-potential "augmented-plane-wave plus local orbital method" as implemented in the Wien2k code[22]. The spin-polarised calculations were performed for the full $Cc$ crystal structure with 112 atoms in the primitive unit cell, using the lattice parameters and atomic positions determined recently at 90 K [21]. The 3$s$, 3$p$, 3$d$, and 4$s$ orbitals of Fe and the 2$s$ and 2$p$ orbitals of O were treated as valence states, and additional local orbitals were used for Fe $s$ and $p$ states and O $s$ states. Electron exchange and correlation were considered in the generalised gradient approximation (GGA) [23] with additional treatment of on-site Coulomb repulsion using the DFT+U approach [24]. An on-site Coulomb energy of $U = 4.5$ eV and an exchange parameter $J = 0.9$ eV were used for all Fe $d$ states as in previous work [15,16,18]. Brillouin zone integration was performed on a regular mesh of 6×6×2 $k$-points with 21 $k$-points in the irreducible part of the Brillouin zone. Atomic sphere muffin tin radii ($R_{MT}$) of 1.86 bohr and 1.65 bohr were chosen for Fe and O, respectively, and the largest plane-wave vector $K_{max}$ was given by $R_{MT}K_{max} = 8$. Spin-orbit coupling was not considered.

The calculated electronic density of states for magnetite around the Fermi energy is shown in Fig. 1. Gaps in both the spin-up and spin-down channels are observed, consistent with the insulating nature of the Verwey phase, and the estimated bandgap in the down-spin states is 500 meV which is somewhat larger than experimental values of ~100 meV from spectroscopic measurements [25]. The narrow down-spin band just below the Fermi level consists almost entirely of $B$ atom $d$-states corresponding to the 'extra' electrons. The spatial distribution of this electron density is of particular interest in relation to the Verwey distortion, and we refer to the states at energies between -460 and 0 meV as the '$B$:$d\downarrow$' band hereafter. $B$elow another 435 meV gap lie broad bands corresponding to the $A$ site (down-spin) and $B$ site (up-spin) $Fe^{3+}$ $3d^5$ states, plus oxygen 2$s$ and 2$p$ contributions.

**Electronic Orders**

To verify the charge ordering deduced from analysis of Fe-O distances in ref. 21, both the total charge density and the $B$:$d\downarrow$ band densities were integrated within the Fe atomic spheres. These are plotted against the structural bond valence sum (BVS) estimates of formal Fe charge for the 16 symmetry independent $B$ sites in the $Cc$ magnetite structure in Fig. 2. Both integrated charges correlate with the BVS estimates, as the $Fe^{2+}$-like sites with the lowest eight BVS values have the highest eight integrated charges, and vice versa for the $Fe^{3+}$-like sites. The $B{:}d\downarrow$ charges cover a range of $0.33e$ (where $e$ is the electron charge) but the range in total charge is smaller ($0.15e$) showing that redistribution of charge from the majority spin $B{:}d$ and oxygen $2s$ and $2p$ bands tends to smear out the charge differences. Neither the integrated charges nor the BVS distributions have the bimodal character expected for an ideal $Fe^{2+}/Fe^{3+}$ charge ordering, and this evidences the charge redistribution within trimerons (inset of Fig. 2) discussed later.

Orbital ordering in the $Cc$ magnetite structure is apparent from the electron density isosurface [26] for the $B{:}d\downarrow$ band states shown in Fig. 3. The narrowness of the band is reflected in the atomic $3d$-orbital-like nature of the densities at the $B$ sites. $B{:}d\downarrow$ density with $t_{2g}$ symmetry in the axis system of the local $BO_6$ octahedra is observed at each $Fe^{2+}$ site, demonstrating that a well-defined orbital order is present [18]. Significant $B{:}d\downarrow$ density is also apparent at the six $Fe^{3+}$-like sites that participate in trimeron bonding with the $Fe^{2+}$ sites, and this shows various $3d$-orbital symmetries according to the number of trimerons (from one to three) terminating at the site. The two $Fe^{3+}$-like sites that do not participate in trimerons have the lowest integrated charges and highest BVS's in Fig. 2 and are the two atoms that show very small $B{:}d\downarrow$ isosurfaces in Fig. 3, confirming that they are the closest to ideal $Fe^{3+}$ states. The observation of anomalously short $Fe^{2+}$-$Fe^{3+}$ distances perpendicular to the local orbital order axes at $Fe^{2+}$ sites in the 90 K $Cc$ magnetite structure [21] was taken as evidence for the formation of linear 'trimeron' units of three coupled $B$ sites in which a minority-spin $t_{2g}$ electron is delocalised from a central $Fe^{2+}$ ion onto two neighbours resulting in weak bonding interactions (see Fig. 2 inset). This is supported by the substantial $B{:}d\downarrow$ density observed at the $Fe^{3+}$-like sites that participate in trimerons in Fig. 3.

The ~3 Å $B$-$B$ connections comprise infinite linear chains parallel to six directions in the magnetite structure. Each $B$ site lies at the intersection of three chains. The $B{:}d\downarrow$ electron density variations in the repeat sequences for all distinct chains are shown in Fig. 4a. Enhanced electron densities at the trimeron termini are observed, consistent with the trimeron model in the inset to Fig. 2 where $B{:}d\downarrow$ electron density is transferred from the central $Fe^{2+}$ ion to the two terminal neighbors. This is quantified using values of the $B{:}d\downarrow$ electron density maxima at the lobes lying ~0.3 Å on both sides of the nucleus in the chain directions. The average $B{:}d\downarrow$ electron density at the maxima of the eight $Fe^{2+}$ trimeron centres in Fig. 4a (between the two T symbols) is 3.47 eÅ$^{-3}$. This high value is in keeping with the orbital order in these directions. The corresponding average for the 16 terminal trimeron lobes (15 of which are from $Fe^{3+}$-type sites) is 0.45 eÅ$^{-3}$, whereas the average for the 24 other lobes (from 15 $Fe^{2+}$-like and 9 $Fe^{3+}$-like sites, traversed in non-trimeron directions) is 0.07 eÅ$^{-3}$. The disparity between the latter two averages would not be expected in a simple charge and orbital ordered model, and evidences the charge transfer within trimerons that smears out the first approximation $Fe^{2+}/Fe^{3+}$ charge order.

Charge transfer also results in a relatively large $B:d\downarrow$ electron density throughout the line connecting the three $B$ sites in each trimeron. This is seen by plotting the $B:d\downarrow$ electron density variations along the central sections (~1 Å in extent) of the straight lines connecting nearest-neighbour pairs of $B$ sites, outside the $R_{MT}$ radii of the two Fe atoms. Near-symmetric pairs of high $B:d\downarrow$ density central sections are observed for most of the predicted $B$-$B$ trimeron pairs in Fig. 4b, and very low density is observed elsewhere. However, a more complex electron localisation is observed along one of the [111] chains (second row in Figs. 4a and 4b) as discussed below. These trimeron orderings are equivalent to static charge density waves of four- or eight- $B$ atom periodicities in the various chain directions.

$B$-site chains parallel to the [100] and [010] directions have four-atom repeat sequences. One [100] chain (consisting of $B$1-- sites derived from the $B$1 position in the previously reported P2/c subcell [7]; shown lowest left in Figs. 4a and 4b) is unique to the entire low temperature magnetite structure as it contains no trimerons. As noted in ref. 21, this is the only chain to contain only $Fe^{2+}$-like states, however, orbital ordering results in all their trimerons being directed along other chain directions. $B$-site chains in the <111> type directions have repeat sequences of eight atoms. Each chain intersects a symmetry-generated equivalent of itself at just one, $B$1-- type, site (e.g. the top panels of Figs. 4a and 4b show that equivalent chains parallel to [111] and [1̄11] intersect only at $B$1$B$1 sites) where trimerons are centred in all four cases. One of these chains contains an additional trimeron, (the $B$42 site in the second row in Figs. 4a and 4b) making this the only chain in which unequal, 3 and 5 atom, trimeron spacings occur. This perturbs the $B:d\downarrow$ density especially for the $B$1$B$2 trimeron where the central section density is spread almost symmetrically over four successive $B$ atoms. This could be viewed as a bond-dimer localised charge, spread over the adjacent two neighbours in the chain, but no other evidence for bond-dimer order is observed in the *Cc* structure, and the central section $B$:$d\downarrow$ densities are consistent with trimeron order in all other cases.

The 16 *B-B* contacts identified as being within trimerons are seen to be those with the highest average $B$:$d\downarrow$ central section densities in Fig. 4c. Although 14 of these *B-B* distances are short, high $B$:$d\downarrow$ density is also observed in the central sections of the two long predicted trimeron *B-B* distances, demonstrating that $B$:$d\downarrow$ electron density is not a simple function of *B-B* distance. The lowest trimeron and highest non-trimeron $B$:$d\downarrow$ density value points, which are proximate at $D_{BB}$ = 3.012 and 3.003 Å on Fig. 4c, correspond to the two outer *B-B* distances in the four-atom $B1B2$ $B$:$d\downarrow$ distribution noted above and which represents the greatest perturbation of the trimeron order.

The trimerons observed in the low temperature magnetite structure may be regarded as 'orbital molecules' – locally-coupled orbital states on two or more metal ions within an orbitally-ordered solid. The magnetite trimerons are one-electron quasiparticles with effective spin-1/2. Previously reported species that may be classified as orbital molecules are polyelectronic spin-singlet species, usually two-electron dimers, e.g. in $CuIr_2S_4$ [27] and $MgTi_2O_4$,[28] which can show remarkable arrangements such as the spontaneous chirality of helical structures in $MgTi_2O_4$. However, more complex species like heptameric, 18-electron, spin-singlet $V_7^{17+}$ clusters in $AlV_2O_4$ have also been reported.[29] Taken together, these reports demonstrate that orbital molecules may be identified as a class of quantum electronic states formed through orbital order in structures such as spinel where direct metal-metal interactions are significant. The structure of the Verwey phase of magnetite demonstrates that complex electronic orders can emerge from the self-organisation of orbital molecules in solids, analogous to the formation of complex molecules from atoms in conventional matter.

## Conclusions

In conclusion, we have performed DFT+U calculations of the electronic structure of magnetite using the low temperature *Cc* structure recently determined by X-ray crystallography [21]. A band gap of ~0.5 eV is calculated. The narrow minority spin band just below the Fermi level consisting predominantly of *B* atom *d*-states is of particular interest as it describes the 'extra' electrons that were predicted to be localised by Verwey [1] and subsequent authors. The spatial distribution of this $B$:$d\downarrow$

electron density is consistent with charge ordering and a well-defined orbital ordering at the $Fe^{2+}$-like sites. However, significant transfer of $B$:$d\downarrow$ density to most of the $Fe^{3+}$-like sites is also observed, and the variations in $B$:$d\downarrow$ density between neighboring $B$-$B$ pairs supports a trimeron ordering description in which 'extra' electrons are localised within linear three $B$-atom units. The trimeron units are evenly distributed along most of the infinite $B$-site chains, but an irregular spacing distorts the $B$:$d\downarrow$ distribution in one case. Hence, the overall electronic order in the Verwey structure approximates to a $Fe^{2+}$/$Fe^{3+}$ charge and orbital model but with significant discrepancies that are usefully described as trimeron order. Trimerons are one of several known examples of orbital molecules, which can self-organisation to generate complex electronic arrangements in orbitally-ordered solids.

We acknowledge support from EPSRC, STFC, and the Leverhulme Trust. This work used resources provided by the Edinburgh Compute and Data Facility (ECDF, www.ecdf.ed.ac.uk); the ECDF is partially supported by the eDIKT initiative (www.edikt.org.uk).

*Corresponding author; j.p.attfield@ed.ac.uk

1  E. J.W. Verwey, Nature (London) **144**, 327 (1939).

2  F. Walz, J. Phys. Condens. Matter **14**, R285 (2002).

3  J. Yoshida, and S. Iida, J. Phys. Soc. Jpn. **42**, 230 (1977).

4  M. Iizumi, et al, Acta Crystallogr B **38**, 2121 (1982).

5  J.P. Attfield, Solid State Sci. **8**, 861 (2006).

6  J.P. Wright, J.P. Attfield, and P.G. Radaelli, Phys. Rev. Lett. **87,** 266401 (2001).

7  J.P. Wright, J.P. Attfield, and P.G. Radaelli, Phys. Rev. B **66,** 214422 (2002).

8  J. Blasco, J. Garcia, and G. Subias, Phys. Rev. B **83,** 104105 (2011).

9  R.J. Goff, J.P. Wright, J.P. Attfield, and P.G. Radaelli, J. Phys.: Condens. Matter **17**, 7633 (2005).

10  E. Nazarenko, et al, Phys. Rev. Lett. **97**, 056403 (2006).

11  Y. Joly, et al, Phys. Rev. B **78,** 134110 (2008).

12  S.R. Bland, et al, J. Phys.: Condens. Matter **21**, 485601 (2009).

13  J.E. Lorenzo, et al, Phys. Rev. Lett. **101**, 226401 (2008).

14  H. Seo, M. Ogata, and H. Fukuyama, Phys. Rev. B **65**, 085107 (2002).

15  H.T. Jeng, G.Y. Guo, and D.J. Huang, Phys. Rev. Lett. **93**, 156403 (2004).

16  H.T. Jeng, G.Y. Guo, and D.J. Huang, Phys. Rev. B **74**, 195115 (2006).

17  J. van den Brink, and D.I. Khomskii, J. Phys.: Condens. Matter **20**, 434217 (2008).

18  K. Yamauchi, T. Fukushima, and S. Picozzi, Phys. Rev. B **79**, 212404 (2009).

19  F. Zhou, and G. Ceder, Phys. Rev. B **81**, 205113 (2010).

20  T. Fukushima, K. Yamauchi, and S. Picozzi, J. Phys. Soc. Jpn. **80**, 014709 (2011).

21  M.S. Senn, J.P. Wright and J.P. Attfield, Nature (London) **481**, 173 (2012).

22  P. Blaha, K. Schwarz, G. K. H. Madsen, D. Kvasnicka, and J. Luitz, *Wien2k, An Augmented Plane Wave + Local Orbitals Program for Calculating Crystal Properties* (K. Schwarz, Techn. Universität Wien, Austria, 2001).

23  J. P. Perdew, K. Burke, and M. Ernzerhof, Phys. Rev. Lett. **77**, 3865 (1996).

24  V. I. Anisimov, I. V. Solovyev, M. A. Korotin, M. T. Czyżyk, G. A. Sawatzky, Phys. Rev. B **48**, 16929 (1993).

25  L. V. Gasparov, D. B. Tanner, D. B. Romero, H. Berger, G. Margaritondo, and L. Forro, Phys. Rev. B **62**, 7939 (2000).

26  K. Momma and F. Izumi, J. Appl. Crystallogr. **44**, 1272 (2011).

27 P. G.  Radaelli et al, Nature (London) **416**, 155 (2002).

28 M. Schmidt et al, Phys. Rev. Lett. **92**, 056402 (2004).

29 Y. Horibe et al Phys. Rev. Lett. **96,** 086406 (2006).

**Fig. 1** (Color online) Spin-dependent electronic density-of-states for the *Cc* Verwey structure of magnetite calculated from experimental structural parameters in ref. 21. Energies are shown relative to the Fermi level which lies at the top of the *B:d↓* band. Contributions from the *A* (tetrahedrally-coordinated Fe), *B* (octahedrally-coordinated Fe) and O atoms are shown below the total density. Upper/lower sections correspond to minority (down)/majority (up) spin states in all plots.

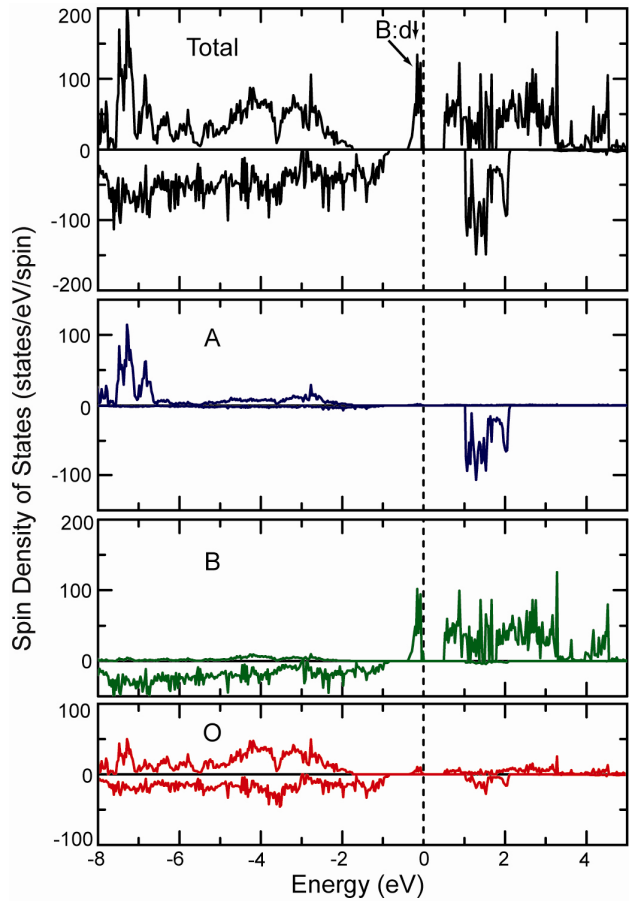

**Fig. 2** (Color online) Integrated total charges, $Q_{Tot}$, and charges from the $B$:$d\downarrow$ band, $Q_{B:d\downarrow}$ for the 16 independent $B$ iron sites in the $Cc$ magnetite structure, plotted against the structural bond valence sum (BVS) estimates of formal charge state in ref. 21. Symbols with dark/light fill correspond to the identified $Fe^{2+}$/$Fe^{3+}$-like states. Inset shows the idealised local structure of a trimeron. A minority spin $d$-electron is localized at one of the $t_{2g}$ orbitals on the central $B$ atom (an $Fe^{2+}$ state). This orbital order elongates the four Fe-O bonds in the plane of the occupied $t_{2g}$ orbital, and weak bonding interactions transfer electron density into coplanar $t_{2g}$ orbitals at two of the six neighbouring $B$ sites, which tends to contract the distances to these two atoms, and diminishes the charge difference between $Fe^{2+}$- and $Fe^{3+}$- like states.

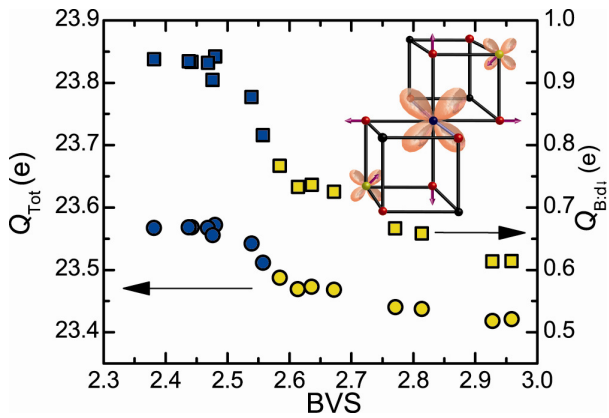

**Fig. 3** (Color online) The *Cc* unit cell of magnetite with *B* site $Fe^{2+}/Fe^{3+}$-like states drawn as dark/light spheres, and trimeron connections between *B* sites shown as proposed in ref. 21. The isosurface for the B:$d\downarrow$ band states at an electron density of 0.1 e/Å$^3$ is shown in the lower part of the unit cell for the 16 independent B sites, labelled as in ref. 21.

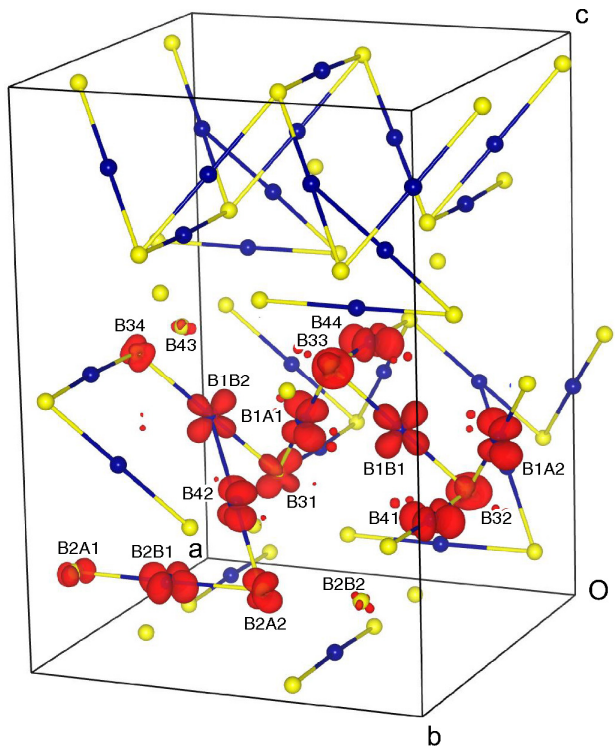

Fig. 4 (Color online) Plots of calculated $B:d\downarrow$ electron density ($\rho$) for Fe atoms in the $Cc$ magnetite structure. (a) and (b) display variations in $\rho$ along the $B$-$B$ vectors for all of the independent linear $B$ chain repeat sequences, with their lattice directions shown. $B$ site labels are the same as in Fig. 3 and ref. 21, and show how the sites are related to those in an earlier $P2/c$ subcell model [7]. The eight consecutive $B$-$B$ pairs assigned to trimerons are indicated by 'T' symbols. (a) $B:d\downarrow$ density variations for $\rho > 10^{-3}$ eÅ$^{-3}$ on a log scale, plotted against cumulative $B$-to-$B$ chain distance $D$. All plots have the same scales. (b) $\rho$ variations in the central sections between neighboring $B$ site Fe atoms. Each central distance scale $D_I$ is along the straight line within a $B$-$B$ pair, starting 1 Å from the first atom center and ending 1 Å from the second. All plots use the same scales. (c) Plot of the average central section electron densities from the 48 panels in (b) against the $B$-$B$ distances. The symbols show the idealised charge states for the two $B$ sites, and closed/open symbols indicate expected trimeron/non-trimeron $B$-$B$ pairs.

(a)

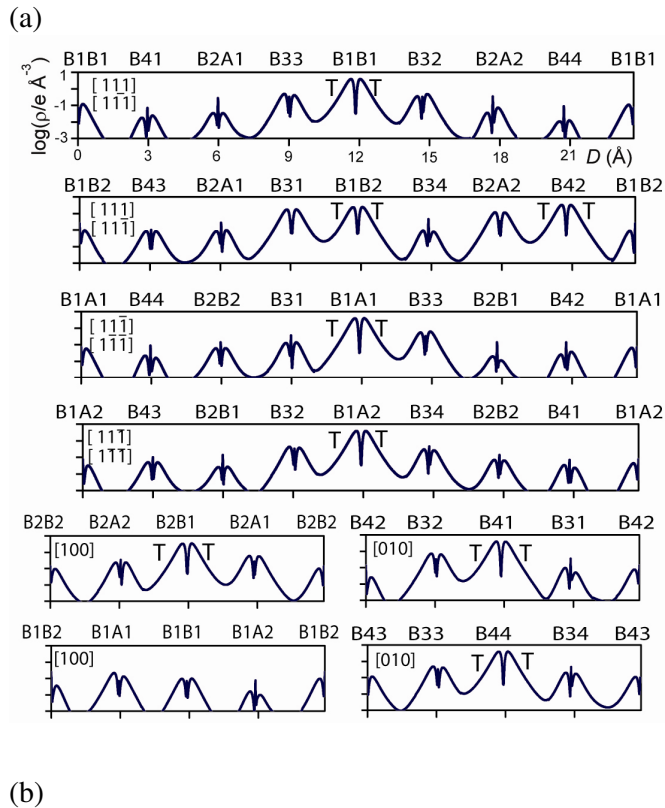

(b)

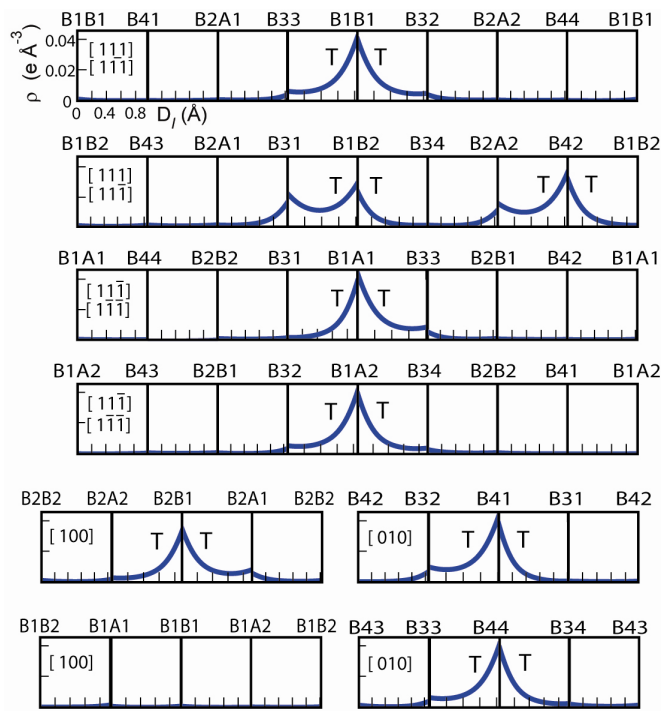

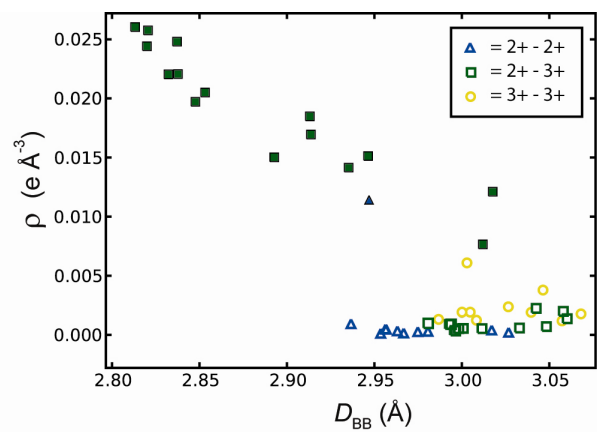

(c)